\documentclass[12pt]{article}

\newcommand{\alphab}{\mbox{\boldmath $\alpha$}}
\newcommand{\ab}{\mbox{\boldmath $a$}}
\newcommand{\xib}{\mbox{\boldmath $\xi$}}
\newcommand{\bb}{\mbox{\boldmath $b$}}
\newcommand{\tv}{\tilde{V}}
\newcommand{\lambdab}{\mbox{\boldmath $\lambda$}}
\newcommand{\tx}{\tilde{x}}
\newcommand{\tphi}{\tilde{\phi}}
\newcommand{\dpp}{\dot{P}}
\def\sech{\mathop{\rm sech}\nolimits}
\def\case#1#2{{\textstyle{#1\over #2}}}
\def\csch{\mathop{\rm csch}\nolimits}

\title{
Deformed shape invariance and exactly solvable Hamiltonians with position-dependent
effective mass}
\author{B Bagchi$^a$, A Banerjee$^a$, C Quesne$^{b}$, V M Tkachuk$^c$\\ 
{\small
$^a$ Department of Applied Mathematics, University of Calcutta,} \\ {\small 92 Acharya
Prafulla Chandra Road, Kolkata 700 009, India}\\ 
{\small $^b$ Physique Nucl\'eaire
Th\'eorique et Physique Math\'ematique,  Universit\'e Libre de Bruxelles,} \\ 
{\small Campus de la Plaine CP229, Boulevard~du Triomphe, B-1050
Brussels, Belgium}\\
{\small $^c$ Ivan Franko Lviv National University, Chair of Theoretical
Physics,}\\
{\small 12, Drahomanov Street, Lviv UA-79005, Ukraine}\\
{\small E-mail: bbagchi123@rediffmail.com, cquesne@ulb.ac.be,
tkachuk@ktf.franko.lviv.ua}}
\date{ }
\begin{document}
\baselineskip=22pt plus 1pt minus 1pt
\maketitle

\begin{abstract} 
Known shape-invariant potentials  for the constant-mass Schr\"odinger equation are
taken as effective potentials in a position-dependent effective mass (PDEM) one. The
corresponding shape-invariance condition turns out to be deformed. Its solvability
imposes the form of both the deformed superpotential and the PDEM. A lot of new
exactly solvable potentials associated with a PDEM background are generated in this way.
A novel and important condition restricting the existence of bound states whenever the
PDEM vanishes at an end point of the interval is identified. In some cases, the
bound-state spectrum results from a smooth deformation of that of the conventional
shape-invariant potential used in the construction. In others, one observes a generation or
suppression of bound states, depending on the mass-parameter values. The corresponding
wavefunctions are given in terms of some deformed classical orthogonal polynomials. 
\end{abstract}


%
%
\newpage
\section{Introduction}

There has been a growing interest in studying
position-dependent-effective-mass (PDEM) quantum Hamiltonians due to their relevance
in describing the dynamics of electrons in many condensed-matter systems, such as
compositionally graded crystals~\cite{geller}, quantum dots~\cite{serra} and liquid
crystals~\cite{barranco}. The PDEM concept has been considered in the
energy-density functional approach to the quantum many-body problem in the context of
nonlocal terms of the accompanying potential and applied to nuclei~\cite{ring}, quantum
liquids~\cite{arias} and metal clusters~\cite{puente}, for instance. Some other theoretical
advances include the derivation of the underlying electron Hamiltonian from instantaneous
Galilean invariance~\cite{levy} and the calculation of Green's function for step and
rectangular-barrier potentials and masses~\cite{chetouani} by implementing path-integral
techniques~\cite{yung}.\par
%
%
Many recent developments have aimed at deriving exact solutions of the PDEM
Schr\"odinger equation (SE)~\cite{dekar, milanovic, plastino, dutra, roy, koc, alhaidari,
gonul, cq04, bagchi, yu}. They have been achieved by extending some well-known
methods used to generate exactly solvable (ES), quasi-ES or conditionally ES potentials.
Such methods include point canonical transformations~\cite{bhatta}, Lie algebraic
methods~\cite{alhassid}, as well as supersymmetric quantum mechanical (SUSYQM) and
shape-invariance (SI) techniques~\cite{gendenshtein, cooper}.\par
%
%
In a recent paper~\cite{cq04}, Quesne and Tkachuk have pointed out certain intimate
connections between the PDEM SE  and the constant-mass SE based on deformed
canonical commutation relations (see also~\cite{mizrahi} for a treatment on the classical
aspect). Their study exploits the existence of a specific relation
between the PDEM and the deforming function appearing in the generalized canonical
commutation relations. As a consequence of this relation, the potential in the deformed SE
may be considered as the effective potential in the PDEM one, taking into account the
interplay of the initial potential and the ambiguity-parameter-dependent contribution of
the kinetic energy term coming from the momentum and mass-operator
noncommutativity.\par
%
%
In the approach of~\cite{cq04}, solving a PDEM SE for a specific choice of the mass
function amounts to considering some deformed SI condition in a SUSYQM framework.
This relates the PDEM formalism to an important branch of SUSYQM, whose development
dates back to that of quantum groups and $q$-algebras and which has produced a lot of
interesting results (see, e.g., \cite{spiridonov, khare, sukhatme, loutsenko}).\par
%
%
The procedure proposed in~\cite{cq04} has been illustrated by considering the case of
the three-dimensional Coulomb problem bound-state energy spectrum. This example has
revealed two interesting features. First, the ambiguity parameters have been shown to
essentially lead to reparametrizing the Coulomb potential without changing its shape.
Second, a drastic effect of the mass environment on the energy spectrum has been
uncovered in the sense that the infinite bound-state spectrum of the constant-mass case
is converted into a finite one.\par
%
%
Both of these results strongly contrast with those of most constructions of solvable
PDEM SE's, where the potential gets mass deformed in a rather complicated way while the
spectrum remains the same as in the constant-mass case. One notable exception to this
general observation comes from a recent analysis of the free-particle problem, where the
presence of a suitable mass environment generates an infinite number of bound
states~\cite{bagchi}.\par
%
%
In this paper, our primary concern is to extend the procedure of~\cite{cq04} to those
one-dimensional potentials that are SI under parameter translation~\cite{cooper}. We
actually plan to show that under some suitable assumptions on the corresponding
superpotential, one may find a PDEM or, equivalently, a deforming function, for
which the deformed SI condition remains solvable, thereby leading to exact results for the
bound-state spectrum and the corresponding wavefunctions of the associated SE's,
provided the latter satisfy some appropriate conditions. Our secondary purposes consist in
studying the interplay of the two contributions to the effective potential and the
generation of the corresponding ES PDEM potential, as well as in determining whether the
associated mass function has a dramatic or only smooth effect on the bound-state
spectrum.\par
%
%
In section~2, the general procedure for solving PDEM SE's through the use of a deformed
SI condition is reviewed. In section~3, various classes of superpotentials are identified.
The method is then illustrated in section~4 by considering some simple examples. The
general results, listed in the appendix, are commented in section~5. Finally, section~6
contains the conclusion.\par
%
%
\section{General procedure}

One of the well-known problems of the PDEM SE consists in the momentum and
mass-operator noncommutativity and the resultant ordering ambiguity in the kinetic
energy term (see, e.g., \cite{levy, morrow, ribeiro, cavalcante}). To cope with this
difficulty, it is advantageous to use the von Roos general two-parameter form of the
effective-mass kinetic energy operator~\cite{vonroos}, which has an inbuilt Hermiticity
and contains other plausible forms as special cases.\par
%
%
In units wherein $\hbar = 2 m_0 = 1$, we may therefore write the PDEM SE as
\begin{eqnarray}
  &&\Biggl[- \frac{1}{2} \left(M^{\xi'}(\alphab; x) \frac{d}{dx} M^{\eta'}(\alphab; x)
        \frac{d}{dx} M^{\zeta'}(\alphab; x) + M^{\zeta'}(\alphab; x) \frac{d}{dx}
        M^{\eta'}(\alphab; x) \frac{d}{dx} M^{\xi'}(\alphab; x)\right) \nonumber \\ 
  && + V(\ab; x)\Biggr] \psi(x) =E \psi(x)  \label{eq:PDEM-SE}
\end{eqnarray}
where $M(\alphab; x)$ is the dimensionless form of the mass function $m(\alphab; x) =
m_0 M(\alphab; x)$, $\alphab$ and $\ab$ denote two sets of parameters, and the von
Roos ambiguity parameters $\xi'$, $\eta'$, $\zeta'$ are constrained by the condition
$\xi' + \eta' + \zeta' = -1$.\par
%
%
On setting 
\begin{equation}
  M(\alphab; x) = \frac{1}{f^2(\alphab; x)} \qquad f(\alphab; x) = 1 + g(\alphab; x)
  \label{eq:M-f}
\end{equation}
where $f(\alphab; x)$ is some positive-definite function and $g(\alphab; x) = 0$
corresponds to the constant-mass case, equation (\ref{eq:PDEM-SE}) becomes 
\begin{eqnarray}
  &&\Biggl[- \frac{1}{2} \left(f^{\xi}(\alphab; x) \frac{d}{dx} f^{\eta}(\alphab; x)
        \frac{d}{dx} f^{\zeta}(\alphab; x) + f^{\zeta}(\alphab; x) \frac{d}{dx}
        f^{\eta}(\alphab; x) \frac{d}{dx} f^{\xi}(\alphab; x)\right) \nonumber \\ 
  && + V(\ab; x)\Biggr] \psi(x) =E \psi(x)  \label{eq:PDEM-SE-bis}
\end{eqnarray}
with $\xi + \eta + \zeta = 2$. Among those ambiguity parameter choices that have been
found useful for describing the motion of electrons in compositionally graded crystals, we
may quote those of BenDaniel and Duke (BDD)~\cite{bendaniel} ($\xi = 0$, $\zeta = 0$),
Bastard~\cite{bastard} ($\xi = 2$, $\zeta = 0$), Zhu and Kroemer (ZK)~\cite{zhu} ($\xi
= 1$, $\zeta = 1$) and Li and Kuhn (LK)~\cite{li} ($\xi = 0$, $\zeta = 1$).\par
%
%
We can get rid of the ambiguity parameters $\xi$, $\eta$, $\zeta$ (denoted collectively
by $\xib$) in the kinetic energy term by transferring them to the effective potential
energy of the variable-mass system. Thus using the result
\begin{eqnarray}
  \lefteqn{f^{\xi} \frac{d}{dx} f^{\eta} \frac{d}{dx} f^{\zeta} +
        f^{\zeta} \frac{d}{dx} f^{\eta} \frac{d}{dx} f^{\xi}} \nonumber \\
  & = & 2 \sqrt{f}\, \frac{d}{dx} f \frac{d}{dx} \sqrt{f} - (1 - \xi - \zeta) f f'' - 2 \left(
        \frac{1}{2} - \xi\right) \left(\frac{1}{2} - \zeta\right) f^{\prime2}
\end{eqnarray}
where a prime denotes derivative with respect to $x$ and the positive definiteness of $f$
is explicitly used, equation~(\ref{eq:PDEM-SE-bis}) acquires the form
\begin{equation}
  H \psi(x) \equiv \left[- \left(\sqrt{f(\alphab; x)}\,\frac{d}{dx} \sqrt{f(\alphab; x)}
  \right)^2 + V_{\rm eff}(\bb; x)\right]
  \psi(x) = E \psi(x)  \label{eq:PDEM-SE-ter}
\end{equation}
in which the effective potential 
\begin{equation}
  V_{\rm eff}(\bb; x) = V(\ab; x) + \tv(\alphab, \xib; x)  \label{eq:Veff} 
\end{equation}
contains an additional mass- and ambiguity-parameter-depending term
\begin{equation}
  \tv(\alphab, \xib; x) = \rho f(\alphab; x) f''(\alphab; x) + \sigma f^{\prime2}(\alphab;
  x).  \label{eq:Vtilde} 
\end{equation}
In (\ref{eq:PDEM-SE-ter}) and (\ref{eq:Veff}), the parameters $\bb$ depend on the
whole set of parameters $\ab$, $\alphab$ and $\xib$, while in (\ref{eq:Vtilde}) we have
denoted by $\rho$ and $\sigma$ the following two ambiguity-parameter combinations
\begin{equation}
  \rho = \frac{1}{2} (1 - \xi - \zeta) \qquad \sigma = \left(\frac{1}{2} - \xi\right)
  \left(\frac{1}{2} - \zeta\right). 
\end{equation}
For the special ambiguity-parameter choices referred to hereabove, they take the values
$\rho = \frac{1}{2}$, $\sigma = \frac{1}{4}$ (BDD), $\rho = - \frac{1}{2}$, $\sigma = -
\frac{3}{4}$ (Bastard), $\rho = - \frac{1}{2}$, $\sigma = \frac{1}{4}$ (ZK), or $\rho =
0$, $\sigma = - \frac{1}{4}$ (LK).\par
%
%
The PDEM SE (\ref{eq:PDEM-SE-ter}) may now be reinterpreted  as a deformed SE
\begin{equation}
  H \psi(x) = \left[\pi^2 + V_{\rm eff}(\bb; x)\right] \psi(x) = E \psi(x) 
  \label{eq:def-SE} 
\end{equation}
corresponding to the replacement of the momentum operator $ p = - {\rm i}
\frac{d}{dx}$ by some deformed one
\begin{equation}
  \pi \equiv \sqrt{f(\alphab; x)}\,p \sqrt{f(\alphab; x)} = - {\rm i}
  \sqrt{f(\alphab; x)}\,\frac{d}{dx} \sqrt{f(\alphab; x)}.  \label{eq:pi}
\end{equation}
With this substitution, the conventional commutation relation $[x, p] = \rm i$ is changed
into
\begin{equation}
  [x, \pi] = {\rm i} f(\alphab; x)
\end{equation}
where $f(\alphab; x)$ acts as a deforming function.\par
%
%
In this paper, we plan to take for $V_{\rm eff}(\bb; x)$ some known SI potential. This
means that the initial potential in the PDEM SE (\ref{eq:PDEM-SE-bis}) will then be
determined by inverting (\ref{eq:Veff}) as
\begin{equation}
  V(\ab; x) = V_{\rm eff}(\bb; x) - \tv(\alphab, \xib; x)  \label{eq:V}
\end{equation}
where the parameters $\ab$ now depend on the SI potential parameters $\bb$ and on
$\alphab$, $\xib$.\par
%
%
To solve equation (\ref{eq:def-SE}) (and therefore (\ref{eq:PDEM-SE-bis})), we will show
that for some appropriately chosen deforming function $f(\alphab; x)$, $H$ may be
considered as the first member $H_0 = H$ of a hierarchy of Hamiltonians
\begin{equation}
  H_i = A^+(\alphab, \lambdab_i) A^-(\alphab, \lambdab_i) + \sum_{j=0}^i \epsilon_j
  \qquad i = 0, 1, 2, \ldots  \label{eq:H_i} 
\end{equation}
where the first-order operators
\begin{equation}
  A^{\pm}(\alphab, \lambdab_i) = \mp \sqrt{f(\alphab; x)}\,\frac{d}{dx} \sqrt{f(\alphab;
  x)} + W(\lambdab_i; x)  \label{eq:A}
\end{equation}
satisfy a deformed SI condition
\begin{equation}
  A^-(\alphab, \lambdab_i) A^+(\alphab, \lambdab_i) = A^+(\alphab, \lambdab_{i+1})
  A^-(\alphab, \lambdab_{i+1}) + \epsilon_{i+1} \qquad i = 0, 1, 2, \ldots
  \label{eq:deformed-SI}  
\end{equation}
and $\epsilon_i$, $i=0$, 1, 2,~\ldots, are some constants. It follows from
equation (\ref{eq:deformed-SI}) that we can rewrite $H_{i+1}$ as
\begin{equation}
  H_{i+1} = A^-(\alphab, \lambdab_i) A^+(\alphab, \lambdab_i) + \sum_{j=0}^i 
  \epsilon_j \qquad i=0, 1, 2, \ldots
\end{equation}
so that the Hamiltonians (\ref{eq:H_i}) fulfil intertwining relations
\begin{equation}
  H_i A^+(\alphab, \lambdab_i) = A^+(\alphab, \lambdab_i) H_{i+1} \qquad
  A^-(\alphab, \lambdab_i) H_i = H_{i+1} A^-(\alphab, \lambdab_i)
\end{equation}
similar to those of the undeformed case.\par
%
%
Solving equation (\ref{eq:deformed-SI}) means that it is possible to find a superpotential
$W(\lambdab; x)$, a deforming function $f(\alphab; x)$ and some constants
$\lambdab_i$, $\epsilon_i$, $i=0$, 1, 2,~\ldots, with $\lambdab_0 =
\lambdab$, such that
\begin{equation}
  V_{\rm eff}(\bb; x) = W^2(\lambdab; x) - f(\alphab; x) W'(\lambdab; x) + \epsilon_0
  \label{eq:C1}
\end{equation}
and 
\begin{equation}
  W^2(\lambdab_i; x) + f(\alphab; x) W'(\lambdab_i; x) =  W^2(\lambdab_{i+1}; x) -
  f(\alphab; x) W'(\lambdab_{i+1}; x) + \epsilon_{i+1} \qquad i = 0, 1, 2, \ldots.
  \label{eq:C2}
\end{equation}
As a consequence, the (deformed) SUSY partner $H_1$ of $H$ will be characterized by a
potential
\begin{equation}
  V_{{\rm eff},1}(\bb, \alphab, \lambdab; x) = V_{\rm eff}(\bb; x) + 2 f(\alphab; x)
  W'(\lambdab; x).  \label{eq:partner}
\end{equation}
\par
%
%
To find a solution to equations (\ref{eq:C1}) and (\ref{eq:C2}), we shall be guided by
our  knowledge of the superpotential $W$ in the undeformed case ($f=1$ or
$g=0$)~\cite{cooper}, where the parameters $\lambdab$ are entirely determined by the
potential parameters $\bb$. Our strategy will consist in (i) assuming that the
deformation does not affect the form of $W$ but only brings about a change in its
parameters $\lambdab$ (which will now also depend on $\alphab$), and (ii) choosing
$g(\alphab; x)$ in such a way that in (\ref{eq:C1}) and (\ref{eq:C2}) the function
$g(\alphab;x) W'(\lambdab; x)$ contains the same kind of terms as those already
present in the undeformed case, i.e., $W^2(\lambdab; x)$ and $W'(\lambdab; x)$. In
section~3, we shall put this recipe into practice for general classes of superpotentials and
determine the accompanying deforming function $f(\alphab; x)$, from which the
corresponding PDEM can then be obtained through equation (\ref{eq:M-f}).\par
%
%
It is worth noting that although on solving equation (\ref{eq:C1}), $\lambdab$ will
become a known function of $\bb$ and $\alphab$, it will often prove convenient to keep
it as a (redundant) argument in operators, energies and wavefunctions.\par
%
%
Having found a solution to equations (\ref{eq:C1}) and (\ref{eq:C2}), we can determine
the bound-state energy spectrum and corresponding wavefunctions of $H$ by an
extension of the conventional SUSYQM and SI
procedure~\cite{gendenshtein, cooper}. Thus the energy eigenvalues are given by
\begin{equation}
  E_n(\alphab, \lambdab) = \sum_{i=0}^n \epsilon_i  \label{eq:E}
\end{equation}
while the ground- and excited-state wavefunctions are obtained by solving the first-order
differential equation
\begin{equation}
  A^-(\alphab, \lambdab) \psi_0(\alphab, \lambdab; x) = 0  \label{eq:eq-1}
\end{equation}
and the recursion relation
\begin{equation}
  \psi_{n+1}(\alphab, \lambdab; x) = [E_{n+1}(\alphab, \lambdab) - E_0(\alphab, 
  \lambdab)]^{-1/2} A^+(\alphab, \lambdab) \psi_n(\alphab, \lambdab_1; x) 
  \label{eq:eq-2}
\end{equation}
respectively.\par
%
%
Equations (\ref{eq:E}), (\ref{eq:eq-1}) and (\ref{eq:eq-2}) only provide formal solutions to
equation (\ref{eq:PDEM-SE-ter}) or (\ref{eq:def-SE}). To be physically acceptable, the
bound-state wavefunctions should indeed satisfy two conditions:\par
\noindent (i) As in conventional quantum mechanics, they should be square integrable on
the (finite or infinite) interval of definition of $V_{\rm eff}(\bb; x)$, i.e.,
\begin{equation}
  \int_{x_1}^{x_2} dx\,  |\psi_n(\alphab, \lambdab; x)|^2 < \infty.  \label{eq:wf-C1}
\end{equation}
\noindent (ii) Furthermore, they should ensure the Hermiticity of $H$. For such a
purpose, it is enough to impose that the deformed momentum operator $\pi$, defined in
(\ref{eq:pi}), be Hermitian. This amounts to the condition
\begin{eqnarray}
  \lefteqn{\int_{x_1}^{x_2} dx\, \psi^*(x) \sqrt{f(\alphab; x)} \left(- {\rm i}
       \frac{d}{dx}\right) \sqrt{f(\alphab; x)}\, \phi(x)} \nonumber \\
  & = & \left[\int_{x_1}^{x_2} dx\, \phi^*(x) \sqrt{f(\alphab; x)}
       \left(- {\rm i} \frac{d}{dx}\right) \sqrt{f(\alphab; x)}\, \psi(x)\right]^* 
       \label{eq:hermite}
\end{eqnarray}
for any $\psi(x)$, $\phi(x) \in L^2(x_1, x_2)$. Integrating the left-hand side of
(\ref{eq:hermite}) by parts leads to
\begin{eqnarray}
  \lefteqn{\int_{x_1}^{x_2} dx\, \psi^*(x) \sqrt{f(\alphab; x)} \left(- {\rm i}
       \frac{d}{dx}\right) \sqrt{f(\alphab; x)}\, \phi(x)}\nonumber  \\
  & = & - {\rm i} \psi^*(x) \phi(x) f(\alphab; x)\bigg|_{x_1}^{x_2} + \int_{x_1}^{x_2}
       dx\, \phi(x) \sqrt{f(\alphab; x)} \left({\rm i} \frac{d}{dx}\right) \sqrt{f(\alphab; x)}\,
       \psi^*(x). 
\end{eqnarray}
Comparison with the right-hand side of (\ref{eq:hermite}) then provides us with the
condition $\psi^*(x) \phi(x) f(\alphab; x) \to 0$ for $x \to x_1$ and $x \to x_2$. This
shows that one has to place the restriction
\begin{equation}
  |\psi_n(\alphab, \lambdab; x)|^2 f(\alphab; x) \to 0 \qquad {\rm for\ } x \to x_1 {\rm
  \ and\ } x \to x_2  \label{eq:wf-C2}
\end{equation}
on the allowed bound-state wavefunctions. This condition will be effective whenever
$f(\alphab; x)$ does not go to some finite constant at the end points  of the interval.\par
%
%
The precise range of $n$ values ($n=0$, 1,~\ldots, $n_{\rm max}$ or $n=0$, 1,
2,~\ldots) in equation (\ref{eq:E}) will therefore be determined by the existence of 
corresponding wavefunctions $\psi_n(\alphab, \lambdab; x)$ satisfying both equations
(\ref{eq:wf-C1}) and (\ref{eq:wf-C2}). In terms of the PDEM (\ref{eq:M-f}), the latter
condition translates into
\begin{equation}
  \frac{|\psi_n(\alphab, \lambdab; x)|^2}{\sqrt{M(\alphab; x)}} \to 0 \qquad {\rm for\ } 
  x \to x_1 {\rm \ and\ } x \to x_2  \label{eq:wf-C2bis}
\end{equation}
which should be checked whenever $M(\alphab; x) \to 0$ for $x \to x_1$ or $x \to
x_2$. It should be stressed that although this condition may be present in any PDEM
problem, it has not been noted so far.\par
%
%
On taking (\ref{eq:A}) into account, the solution of equation (\ref{eq:eq-1}) can be
formally obtained in terms of $W$ and $f$. It is given by
\begin{equation}
  \psi_0(\alphab, \lambdab; x) = \frac{N_0(\alphab, \lambdab)}{\sqrt{f(\alphab; x)}}
  \exp\left(- \int^x \frac{W(\lambdab; \tx)}{f(\alphab; \tx)} d\tx\right)  \label{eq:gs-wf} 
\end{equation}
where $N_0(\alphab, \lambdab)$ is some normalization coefficient.\par
%
%
Similarly, the solution of (\ref{eq:eq-2}) can be shown to be
\begin{equation}
  \psi_n(\alphab, \lambdab; x) = \frac{N_n(\alphab, \lambdab)}{\sqrt{f(\alphab; x)}}\,
   \varphi_n(\alphab, \lambdab; x)\exp\left(- \int^x \frac{W(\lambdab_n; \tx)}
  {f(\alphab; \tx)} d\tx\right)  \label{eq:exc-wf} 
\end{equation}
where $\varphi_n(\alphab, \lambdab; x)$ fulfils the equation
\begin{equation}
  \varphi_{n+1}(\alphab, \lambdab; x) = - f(\alphab; x) \varphi'_n(\alphab, \lambdab_1;
  x) + [W(\lambdab_{n+1}; x) + W(\lambdab; x)] \varphi_n(\alphab, \lambdab_1; x)
  \label{eq:phi-eq}
\end{equation}
with $\varphi_0(\alphab, \lambdab; x) = 1$, and the normalization coefficient
$N_n(\alphab, \lambdab)$ satisfies the recursion relation
\begin{equation}
  N_{n+1}(\alphab, \lambdab) = [E_{n+1}(\alphab, \lambdab) - E_0(\alphab, \lambdab)]
  ^{-1/2} N_n(\alphab, \lambdab_1).
\end{equation}
\par
%
%
\section{Classes of superpotentials and their accompanying deforming function}
\setcounter{equation}{0}

In this section, we plan to consider several classes of superpotentials, which in the next
sections will prove to include all the SI potentials considered in table 4.1
of~\cite{cooper}, as well as their special cases. For each class, we shall determine the
general form of the accompanying deforming function. From the expressions obtained for
$W$ and $f$, we shall then deduce some consequences regarding the ground- and
excited-state wavefunction explicit form.\par
%
%
\subsection{Classes of superpotentials}

Let $\phi(x)$ be some parameter-independent function of $x$ and $\lambdab$ denote a
single parameter (for class 0) or a set of two parameters $\lambda$, $\mu$ (for classes
1, 2 and 3).\par
%
%
\medskip
\noindent
{\sl Class 0}
\medskip
\par
%
%
The simplest choice of superpotential is a single-parameter one of the type
\begin{equation}
  W(\lambdab; x) = \lambda \phi(x).  \label{eq:class0-1}
\end{equation}
Conditions (\ref{eq:C1}) and (\ref{eq:C2}) then contain two parameters to be
determined, namely $\lambda$, $\epsilon_0$ and $\lambda_{i+1}$, $\epsilon_{i+1}$,
respectively.\par
%
%
In the undeformed case, apart from a constant term, $\epsilon_0$ or $\epsilon_{i+1}$,
they include the functions $W^2$ and $W'$. Since $W^2$ is proportional to $\phi^2$
and we need two equations to calculate the couple of undetermined parameters,
equations (\ref{eq:C1}) and (\ref{eq:C2}) solvability imposes that $W' = \lambda \phi'$
be a linear combination of $\phi^2$ and a constant. In other words, there must exist
some numerical (i.e., parameter-independent) constants $A$ and $B$ such that
\begin{equation}
  \phi'(x) = A \phi^2(x) + B.  \label{eq:class0-2}
\end{equation}
\par
%
%
In the deformed case, equations (\ref{eq:C1}) and (\ref{eq:C2}) contain in addition a
term $W' g = \lambda \phi' g$. If this term has the same form as the remaining ones, it
will not spoil the equations solvability. This amounts to assuming that there exist two
$\alphab$-dependent constants $A'(\alphab)$ and $B'(\alphab)$ such that
\begin{equation}
  \phi'(x) g(\alphab; x) = A'(\alphab) \phi^2(x) + B'(\alphab).  \label{eq:class0-3}
\end{equation}
On combining (\ref{eq:class0-3}) with (\ref{eq:class0-2}), we get
\begin{equation}
  g(\alphab; x) = \frac{A'(\alphab) \phi^2(x) + B'(\alphab)}{A \phi^2(x) + B}
  \label{eq:class0-4}
\end{equation}
which provides us with the general form of the deforming function $f(\alphab; x)$ for
class 0 superpotentials.\par
%
%
\medskip
\noindent
{\sl Class 1}
\medskip
\par
%
%
The most straightforward generalization of (\ref{eq:class0-1}) consists in adding some
nonvanishing parameter $\mu$:
\begin{equation}
  W(\lambdab; x) = \lambda \phi(x) + \mu.  \label{eq:class1-1}  
\end{equation}
Equations (\ref{eq:C1}) and (\ref{eq:C2}) now contain three parameters ($\lambda$,
$\mu$, $\epsilon_0$ or $\lambda_{i+1}$, $\mu_{i+1}$, $\epsilon_{i+1}$) to be
determined, but as a counterpart $W^2$ is also made of three terms proportional to
$\phi^2$, $\phi$ and a constant, respectively. We shall then get three equations to
govern the parameter values both in the undeformed and deformed cases provided
\begin{equation}
  \phi'(x) = A \phi^2(x) + B \phi(x) + C  \label{eq:class1-2}
\end{equation}
and
\begin{equation}
  g(\alphab; x) = \frac{A'(\alphab) \phi^2(x) + B'(\alphab) \phi(x) + C'(\alphab)}{A
  \phi^2(x) + B \phi(x) + C}  \label{eq:class1-3}
\end{equation}
where $A$, $B$, $C$ and $A'(\alphab)$, $B'(\alphab)$, $C'(\alphab)$ are some
numerical and $\alphab$-dependent constants, respectively.\par
%
%
Comparison of equations (\ref{eq:class0-1}), (\ref{eq:class0-2}), (\ref{eq:class0-4}) with
equations (\ref{eq:class1-1}) -- (\ref{eq:class1-3}) shows that class 0 superpotentials
may be considered as special cases of class 1 superpotentials, corresponding to the
simultaneous vanishing of $\mu$, $B$ and $B'(\alphab)$. In the following, we shall
therefore include class 0 into class 1 by assuming that for the latter either $\mu \ne 0$
or $\mu = B = B'(\alphab) = 0$.\par
%
%
\medskip
\noindent
{\sl Class 2}
\medskip
\par
%
%
If we define $W$ as
\begin{equation}
  W(\lambdab; x) = \lambda \phi(x) + \frac{\mu}{\phi(x)}
\end{equation}
where $\lambda$ and $\mu$ are both nonvanishing (otherwise we would get back class
0), $W^2$ again contains three terms proportional to $\phi^2$, $\phi^{-2}$ and a
constant, respectively.\par
%
%
A reasoning similar to that carried out for class 1 superpotentials leads to the following
expressions for $\phi'$ and $g$,
\begin{eqnarray}
  \phi'(x) & = & A \phi^2(x) + B \\
  g(\alphab; x) & = & \frac{A'(\alphab) \phi^2(x) + B'(\alphab)}{A \phi^2(x) + B}
\end{eqnarray}
where $A$, $B$, $A'(\alphab)$ and $B'(\alphab)$ are independent of $x$.\par
%
%
\medskip
\noindent
{\sl Class 3}
\medskip
\par
%
%
On assuming 
\begin{equation}
  W(\lambdab; x) = \frac{\lambda \phi(x) + \mu}{\sqrt{A \phi^2(x) + B}}
\end{equation}
where $\lambda$, $\mu$ are nonvanishing and $A$, $B$ are two numerical nonvanishing
constants (otherwise we would get back one of the previous classes), we obtain after a
simple calculation
\begin{eqnarray}
  W^2(\lambdab; x) & = & \frac{\lambda^2 \phi^2(x) + 2\lambda \mu \phi(x) + \mu^2}
          {A \phi^2(x) + B} \\
  W'(\lambdab; x) & = & \frac{\lambda B - \mu A \phi(x)}{A \phi^2(x) + B}
          \frac{\phi'(x)}{\sqrt{A \phi^2(x) + B}}.
\end{eqnarray}
\par
%
%
Equations (\ref{eq:C1}) and (\ref{eq:C2}) solvability in the undeformed case then
implies that
\begin{equation}
  \phi'(x) = [C \phi(x) + D] \sqrt{A \phi^2(x) + B}
\end{equation}
in terms of two additional numerical constants $C$ and $D$. In the deformed case, the
supplementary term must therefore be given by
\begin{equation}
  \phi'(x) g(\alphab; x) = [C'(\alphab) \phi(x) + D'(\alphab)] \sqrt{A \phi^2(x) + B} 
\end{equation}
where $C'(\alphab)$ and $D'(\alphab)$ depend on the deforming parameters $\alphab$.
Hence we obtain
\begin{equation}
  g(\alphab; x) = \frac{C'(\alphab) \phi(x) + D'(\alphab)}{C \phi(x) + D}.
\end{equation}
\par
%
%
\subsection{Corresponding wavefunctions}

In equation (\ref{eq:gs-wf}), the ground-state wavefunction $\psi_0(\alphab, \lambdab;
x)$ of $H$ is formally given  in terms of the integral of the function $W(\lambdab;
\tx)/f(\alphab; \tx)$. On taking the explicit forms of $W$ and $f$ obtained in section
3.1 into account, it is straightforward to obtain
\begin{eqnarray}
  && \int^x \frac{W(\lambdab; \tx)}{f(\alphab; \tx)}\, d\tx  \nonumber \\
  && = \int^{\phi(x)} \frac{\lambda\tphi + \mu}{[A + A'(\alphab)] \tphi^2 + [B +
       B'(\alphab)] \tphi + C + C'(\alphab)}\, d\tphi \qquad \mbox{\rm for  class 1}
       \nonumber \\
  && = \int^{\phi(x)} \frac{\lambda\tphi^2 + \mu}{\tphi \{[A + A'(\alphab)] \tphi^2 + 
       B + B'(\alphab)\}}\, d\tphi \qquad \mbox{\rm for class 2}  \nonumber \\
  && = \int^{\phi(x)} \frac{\lambda\tphi + \mu}{(A \tphi^2 + B) \{[C + C'(\alphab)]
       \tphi + D + D'(\alphab)\}}\, d\tphi \qquad \mbox{\rm for class 3} 
       \label{eq:gs-int}     
\end{eqnarray}
thus showing that in all three cases the integral can be explicitly carried out by simple
integration techniques as in the undeformed case.\par
%
%
{}Furthermore, it is possible to write the functions $\varphi_n(\alphab, \lambdab; x)$,
entering the general expression (\ref{eq:exc-wf}) of excited-state wavefunctions, in
terms of $n$th-degree polynomials in a new variable $y$, $P_n(\alphab, \lambdab; y)$,
which fulfil some equation deriving from (\ref{eq:phi-eq}). This result generalizes to the
deformed case a well-known property according to which SI potential wavefunctions can
be expressed in terms of some classical orthogonal polynomials~\cite{cooper}.\par
%
%
The precise form of the changes of variable $x \to y$ and of function
$\varphi_n(\alphab, \lambdab;x) \to P_n(\alphab, \lambdab; y)$, as well as the relation
satisfied by $P_n(\alphab, \lambdab; y)$, actually depend on the superpotential class as
listed herebelow:
\begin{itemize}
\item Class 1
\begin{eqnarray}
  \varphi_n(\alphab, \lambdab; x) & = & P_n(\alphab, \lambdab;y) \qquad y = \phi(x) \\
  P_{n+1}(\alphab, \lambdab; y) & = & - \{[A + A'(\alphab)] y^2 + [B + B'(\alphab)] y
       + C + C'(\alphab)\} \dpp_n(\alphab, \lambdab_1; y)  \nonumber \\
  & & \mbox{} + [(\lambda_{n+1} + \lambda) y + \mu_{n+1} + \mu] P_n(\alphab,
        \lambdab_1; y)  \label{eq:class1-P}
\end{eqnarray}
\item Class 2
\begin{eqnarray}
  \varphi_n(\alphab, \lambdab; x) & = & y^{-n/2} P_n(\alphab, \lambdab;y) \qquad y =
       \phi^{-2}(x)  \\
  P_{n+1}(\alphab, \lambdab; y) & = & 2y \{A + A'(\alphab) + [B + B'(\alphab)] y\}
       \dpp_n(\alphab, \lambdab_1; y)  \nonumber \\
  & & \mbox{} + \{\lambda_{n+1} + \lambda - n [A + A'(\alphab)] + [\mu_{n+1} + \mu
       - n (B + B'(\alphab))] y\}  \nonumber \\
  && \mbox{} \times P_n(\alphab, \lambdab_1; y)
\end{eqnarray}
\item Class 3
\begin{eqnarray}
  \varphi_n(\alphab, \lambdab; x) & = & (A y^2 + B)^{-n/2} P_n(\alphab, \lambdab;y)
        \qquad y = \phi(x) \\
  P_{n+1}(\alphab, \lambdab; y) & = & \{[C + C'(\alphab)] y + D + D'(\alphab)\}
        \nonumber \\
  && \mbox{} \times \left[ - (A y^2 + B) \dpp_n(\alphab, \lambdab_1; y) + nAy
        P_n(\alphab, \lambdab_1; y)\right]  \nonumber \\
  & & \mbox{} + [(\lambda_{n+1} + \lambda) y + \mu_{n+1} + \mu] P_n(\alphab,
        \lambdab_1; y).  \label{eq:class3-P}
\end{eqnarray}
\end{itemize}
Here a dot stands for derivative with respect to $y$ and in all cases $P_0(\alphab,
\lambdab; y) \equiv 1$. It should be noted that in (\ref{eq:class3-P}), the linear
combination $- (A y^2 + B) \dpp_n(\alphab, \lambdab_1; y) + nAy P_n(\alphab,
\lambdab_1; y)$ is actually an $n$th-degree polynomial in $y$ because its
$(n+1)$th-degree term vanishes identically.\par
%
%
\section{Some simple examples}
\setcounter{equation}{0}

The purpose of this section is twofold: we demonstrate by means of some simple
examples how our method developed in the previous sections works in practice and then
illustrate the effect of the new restriction (\ref{eq:wf-C2}) or (\ref{eq:wf-C2bis}) placed
by a deformation or PDEM background on an ES potential bound-state spectrum.\par
%
%
\subsection{Particle in a box and trigonometric P\"oschl-Teller potential}

Setting $\mu=0$ and $\phi(x) = \tan x$ in (\ref{eq:class1-1}) leads to the following 
mutually compatible pair of superpotential and function $g(\alpha; x)$:
\begin{equation}
  W(\lambda; x) = \lambda \tan x \qquad g(\alpha; x) = \alpha \sin^2 x \qquad -
  \frac{\pi}{2} \le x \le \frac{\pi}{2} \qquad  -1 < \alpha \ne 0   \label{eq:ex1-W}
\end{equation}
where the range of $\alpha$ restricts the deforming function $f(\alpha; x)$ to be
positive definite in the interval $\left[- \frac{\pi}{2}, \frac{\pi}{2}\right]$, as it should
be. Note that with $\lambda = A$ the above superpotential yields in conventional
quantum mechanics~\cite{sukumar85b} the familiar trigonometric P\"oschl-Teller potential
\begin{equation}
  V_{\rm eff}(A; x) = A(A-1) \sec^2 x \qquad A > 1 \qquad- \frac{\pi}{2} \le x \le
  \frac{\pi}{2}
\end{equation}
having the associated bound-state energies and wavefunctions
characterized by $E_n = (A+n)^2$ and $\psi_n(x) = N_n (\cos x)^A C^{(A)}_n(\sin x)
$, $n=0$, 1, 2,~\ldots~\cite{cq99}. The particle-in-a-box problem being the limiting
case of P\"oschl-Teller for $A \to 1$ corresponds to
\begin{equation}
  V_{\rm eff}(x) = \left\{\begin{array}{ll}
       0 & \mbox{if $ - \frac{\pi}{2} < x < \frac{\pi}{2}$} \\[0.2cm]
       \infty & \mbox{if $ x = \pm \frac{\pi}{2}$}
  \end{array} \right..  \label{eq:box}
\end{equation}
\par
%
%
Let us consider the particle in a box first. Turning to (\ref{eq:C1}) and using
(\ref{eq:ex1-W}), it is straightforward to obtain the solutions
$\lambda = 1 + \alpha$ and $\epsilon_0 = 1 + \alpha$. These imply from (\ref{eq:C2})
$\lambda_i = (i+1) (1 + \alpha)$ and $\epsilon_i = (2i+1) (1 + \alpha)$, $i=0$, 1,
2,~\ldots. Consequently, equation (\ref{eq:E}) furnishes the energy eigenvalues 
\begin{equation}
  E_n(\alpha, \lambda) = (1 + \alpha) (n+1)^2. \label{eq:ex1-E}
\end{equation}
The corresponding wavefunctions are easily obtainable from (\ref{eq:gs-wf}) and
(\ref{eq:exc-wf}), which give the common form
\begin{equation}
  \psi_n(\alpha, \lambda; x) = N_n(\alpha, \lambda) \frac{(\cos x)^{n+1}}{(1 + \alpha
  \sin^2 x)^{(n+2)/2}} P_n(\alpha, \lambda; \tan x) \label{eq:ex1-wf}
\end{equation}
where equations (\ref{eq:gs-int}) -- (\ref{eq:class1-P}) have been used. In
(\ref{eq:ex1-wf}), $P_n(\alpha, \lambda; y)$ satisfies the equation
\begin{equation}
  P_{n+1}(\alpha, \lambda; y) = - [1 + (1+\alpha) y^2] \dpp_n(\alpha, \lambda_1; y) +
  (n+3) (1+\alpha) y P_n(\alpha, \lambda_1; y)  \label{eq:ex1-P}
\end{equation}
with $\lambda = 1 + \alpha$ and $\lambda_1 = 2 + 2\alpha$. For $-1 < \alpha \ne 0$
and any $n=0$, 1, 2,~\ldots, equation (\ref{eq:ex1-wf}) manifestly represents a
square-integrable function in $\left[- \frac{\pi}{2}, \frac{\pi}{2}\right]$. In addition, since
$f\left(\alpha, \pm \frac{\pi}{2}\right) = 1 + \alpha$, condition (\ref{eq:wf-C2}) is also
automatically satisfied. We conclude that in the presence of deformation
(\ref{eq:ex1-W}), the particle-in-a-box problem still has an infinite number of bound states
making up a quadratic spectrum. As can be checked, for $\alpha \to 0$, equations
(\ref{eq:ex1-E}) and (\ref{eq:ex1-wf}) go over to their standard forms because
$P_n(\alpha, \lambda; \tan x) \to \gamma_n \sec^n x\, C^{(1)}_n(\sin x)$.\par
%
%
When translating this property into the PDEM language, we are led to a new ES SE,
corresponding to the mass function given in (\ref{eq:M-f}), (\ref{eq:ex1-W}), and to the
potential (\ref{eq:V}), for which
\begin{equation}
  \tv(\alpha, \rho, \sigma; x) = - (\rho + \sigma) \alpha^2 \cos^2 2x + \rho \alpha
  (2+\alpha) \cos 2x + \sigma \alpha^2 \qquad - \frac{\pi}{2} \le x \le \frac{\pi}{2}. 
  \label{eq:ex1-tV}
\end{equation}
It is worth noting that for the LK choice of ambiguity parameters, the latter
expression assumes a very simple form, namely
\begin{equation}
  \tv(\alpha; x) = - \case{1}{4} \alpha^2 \sin^2 2x \qquad - \frac{\pi}{2} \le x \le
\frac{\pi}{2}.
\end{equation}
\par
%
%
What has been done for the particle-in-a-box problem can be easily
extended to the trigonometric P\"oschl-Teller potential. We skip the details, which are
lengthy but straightforward, and give the final form of the deformed energy levels and the
associated wavefunctions
\begin{eqnarray}
  E_n(\alpha, \lambda) & = & (\lambda+n)^2 - \alpha (\lambda-n^2)
        = \left[\case{1}{2}(\Delta+1) + n\right]^2 + \alpha n(n+1) - \case{1}{4}
        \alpha^2  \label{eq:ex2-E}\\ 
  \psi_n(\alpha, \lambda; x) & = & N_n(\alpha, \lambda) (\cos x)^{\frac{\lambda}
        {1+\alpha} + n} (1 + \alpha \sin^2 x)^{- \frac{1}{2}\left(\frac{\lambda}{1+\alpha}
         +n+1\right)} P_n(\alpha, \lambda; \tan x) \label{eq:ex2-wf}
\end{eqnarray}
where $P_n(\alpha, \lambda; y)$ satisfies the equation
\begin{equation}
  P_{n+1}(\alpha, \lambda; y) = - [1 + (1+\alpha) y^2] \dpp_n(\alpha, \lambda_1;
         y) + [2\lambda + (n+1) (1+\alpha)] y P_n(\alpha, \lambda_1; y).
\end{equation}
All functions $\psi_n(\alpha, \lambda; x)$, $n=0$, 1, 2,~\ldots, satisfy both conditions
(\ref{eq:wf-C1}) and (\ref{eq:wf-C2}) again. In the limit $\alpha \to 0$, the standard
results of the P\"oschl-Teller are recovered.\par
%
%
\subsection{Free particle and hyperbolic P\"oschl-Teller potential}

It is worth comparing what happens for the trigonometric superpotential (\ref{eq:ex1-W})
with the case of its hyperbolic counterpart
\begin{equation}
  W(\lambda; x) = \lambda \tanh x \qquad g(\alpha; x) = \alpha \sinh^2 x \qquad 0 <
  \alpha < 1 \label{eq:ex3-W}
\end{equation}
where the range of $\alpha$ provides us with a positive-definite deforming function
$f(\alpha; x)$ again.\par
%
%
In conventional SUSYQM, the above superpotential has been considered in connection with
the attractive or repulsive $\sech^2 x$ potential (often referred to as the hyperbolic
P\"oschl-Teller potential or barrier)~\cite{sukumar85a}, as well as with their limiting case,
namely the free-particle problem~\cite{sukumar85a, kwong}. Let us therefore consider
\begin{equation}
  V_{\rm eff}(A; x) = - A(A+1) \sech^2 x \qquad A > 0
\end{equation}
corresponding to the hyperbolic P\"oschl-Teller potential (and giving the free-particle
problem for $A \to 0$). In the undeformed case, it is known to support a finite number
$n_{\rm max} + 1$ ($A-1 \le n_{\rm max} < A$) of bound states, whose energies and
wavefunctions are given by $E_n = - (A-n)^2$ and $\psi_n(x) = N_n (\sech x)^{A-n}
C^{(A-n+\frac{1}{2})}_n(\tanh x)$, where $n=0$, 1, \ldots, $n_{\rm
max}$~\cite{nieto}. Such results can be derived by SUSYQM and SI
techniques on using the superpotential (\ref{eq:ex3-W}) with $\lambda = A$ and the
factorization energy $\epsilon_0 = - A^2$~\cite{sukumar85a}.\par
%
%
By proceeding as in section~4.1, in the deformed case we obtain
\begin{equation}
  E_n(\alpha, \lambda) = - (\lambda-n)^2 + \alpha (\lambda+n^2) = 
  - \left[\case{1}{2}(\Delta-1) - n\right]^2 + \alpha n(n+1) + \case{1}{4} \alpha^2
  \label{eq:ex3-E}
\end{equation}
and
\begin{equation}
  \psi_n(\alpha, \lambda; x) = N_n(\alpha, \lambda) (\sech x)^{\frac{\lambda}
 {1-\alpha} - n} (1 + \alpha \sinh^2 x)^{\frac{1}{2}\left(\frac{\lambda}{1-\alpha} -
  n - 1\right)} P_n(\alpha, \lambda; \tanh x)  \label{eq:ex3-wf} 
\end{equation}
where $\lambda$ is defined by $\lambda = \case{1}{2}(\alpha - 1 + \Delta)$, $\Delta
\equiv \sqrt{(1-\alpha)^2 + 4A(A+1)}$.  
\par
%
%
{}For any $n=0$, 1, 2,~\ldots, the function (\ref{eq:ex3-wf}) is square integrable
because $\psi_n(\alpha, \lambda; x) \sim e^{-|x|}$ for $x \to \pm \infty$. However, in
the same limits, the deforming function $f(\alpha; x)$, given by (\ref{eq:M-f}) and
(\ref{eq:ex3-W}), behaves as $e^{2|x|}$. Hence condition (\ref{eq:wf-C2}), necessary to
ensure the Hermiticity of $\pi$, cannot be satisfied. From this we infer that with a
deformed function corresponding to (\ref{eq:ex3-W}), the hyperbolic P\"oschl-Teller
potential has no bound state. The same result remains valid for the free-particle problem
and contrasts with what was obtained in~\cite{bagchi} in another context. While this
shows that the result is strictly environment dependent, one must also remember that
in the conventional free-particle problem (see, e.g.,~\cite{davies}) one way to avoid
the divergence is to assume that the particle is confined to a closed and finite
universe. In the context of PDEM a similar philosophy may be adopted with regard to
the preservation of the condition (\ref{eq:wf-C2}).\par
%
%
In conclusion, we have shown how the simple fact of going from trigonometric to
hyperbolic functions in a deformed or PDEM environment may drastically change the
picture as far as an ES potential bound-state spectrum is concerned. In this respect,
the new condition (\ref{eq:wf-C2}) or (\ref{eq:wf-C2bis}), introduced in this paper,
has played an essential role.\par
%
%
\section{Results for shape-invariant potentials}
\setcounter{equation}{0}

The procedure demonstrated on some simple examples in section~4 can be easily
generalized to other shape-invariant potentials. In the appendix, we list some of the results
obtained when taking for $V_{\rm eff}(\bb; x)$ the potentials considered in table 4.1
of~\cite{cooper}. In this respect, two important remarks are in order.\par
%
%
{}First, although the trigonometric P\"oschl-Teller potential of section~4.1 may be
considered as a limiting case of Rosen-Morse I potential when its parameter $B$ goes to
zero (and a change of variable $x \in [0, \pi] \to x' = x - \frac{\pi}{2} \in \left[-
\frac{\pi}{2}, \frac{\pi}{2}\right]$ is performed), the different choices of deforming
function made in section~4.1 and in the appendix produce unrelated results in the
deformed or PDEM context.\par
%
%
Second, three of the potentials listed in~\cite{cooper} are missing from the appendix,
namely Scarf II, Rosen-Morse II and generalized P\"oschl-Teller potentials. The reasons for
their absence are different. For Scarf II potential, it turns out that no nontrivial values of
the parameters may ensure positive definiteness of $f(\alphab; x)$ on the whole real line,
on which the potential is defined. On the other hand, for Rosen-Morse II and generalized
P\"oschl-Teller potentials the resulting square-integrable wavefunctions do not ensure the
Hermiticity of $\pi$, as expressed in condition (\ref{eq:wf-C2}). As a consequence, these
potentials, which both have a finite number of bound states in the undeformed case, do
not support any bound state in the deformed one.\par
%
%
Let us now turn ourselves to the results listed in the appendix. For the SI potentials
$V_{\rm eff}(\bb; x)$ considered there, the potentials $V(\ab; x)$ to be used in the
PDEM equation (\ref{eq:PDEM-SE-bis}) fall into two categories. For the shifted oscillator,
three-dimensional oscillator, Coulomb and Morse potentials, $V(\ab; x)$ has the
same shape as $V_{\rm eff}(\bb; x)$. The only effect of the mass and ambiguity
parameters indeed amounts to a renormalization of the potential parameters and/or an
energy shift $\delta v$. So we obtain
\begin{eqnarray}
  V(\ab; x) & = & \frac{1}{4} \omega^{*2} \left(x - \frac{2b^*}{\omega^*}\right)^2 +
        \delta v \label{eq:SHO}\\
  V(\ab; x) & = & \frac{1}{4} \omega^{*2} x^2 + \frac{l(l+1)}{x^2} + \delta v
        \label{eq:3D-HO}  \\
  V(\ab; x) & = & - \frac{e^2}{x} + \frac{l(l+1)}{x^2} + \delta v \\
  V(\ab; x) & = & B^{*2} e^{-2x} - B^* (2A^*+1) e^{-x} 
\end{eqnarray}
respectively, where, for Morse potential, for instance,
\begin{equation}
  A^* = \frac{1}{2} \left(\frac{B(2A+1) + \rho\alpha}{\sqrt{B^2 - (\rho+\sigma)
  \alpha^2}} - 1\right) \qquad B^* = \sqrt{B^2  - (\rho+\sigma) \alpha^2}. 
\end{equation}
\par
%
%
One may observe strikingly distinct influences of deformation or mass parameters on
bound-state energy spectra. In some cases (shifted oscillator, three-dimensional
oscillator, Scarf I and Rosen-Morse I), the infinite number of bound states of conventional
quantum mechanics remains infinite after the onset of deformation. Similarly, for Morse
potential and for Eckart potential with $\alpha \ne -2$, one keeps a finite number of bound
states. For the Coulomb potential, however, the infinite number of bound states is
converted into a finite one, while for Eckart potential with $\alpha = -2$, the finite
number of bound states becomes infinite.  It is also remarkable that, whenever finite, the
bound-state number becomes dependent on the deforming parameter.\par
%
%
In the appendix, for lack of space we have not shown the explicit form of the
excited-state wavefunctions, in particular that of the polynomials $P_n(\alphab,
\lambdab; y)$\footnote{Detailed results are available from the authors.}. For the same
reason, we have not exhibited the SUSY partners of $V_{\rm eff}(\bb; x)$, which can be
easily determined from equation (\ref{eq:partner}) and reduce to the conventional ones
of~\cite{cooper} in the constant-mass limit.\par
%
%
\section{Conclusion}

In this paper, we have generated a lot of new ES potentials associated with a PDEM
background. For such a purpose, we have considered known SI potentials for the
constant-mass SE as effective potentials in the PDEM one, taking into account the
ambiguity-parameter-dependent contribution coming from the momentum and
mass-operator noncommutativity. The corresponding deformed SI condition solvability
has imposed the general form of both the deformed superpotential and the PDEM. For the
latter, we have then chosen a fairly general particular case and we have found both the
corresponding ES potential and the bound-state energy spectrum and wavefunctions.\par
%
%
The existence of such a spectrum is determined not only by a square-integrability
condition on the wavefunctions as in conventional quantum mechanics, but also by a
Hermiticity condition on the deformed momentum operator. The latter is a new and
important contribution of the present paper. As we have demonstrated on some specific
examples, it may have relevant effects whenever the PDEM vanishes at an
end  point of the interval on which the potential is defined.\par
%
%
We have shown that in some cases the new ES potential has the same shape as the
conventional SI potential used in the construction, but that in others the
ambiguity-parameter-dependent term turns out to change its shape.\par
%
%
{}Furthermore, if in some instances the spectrum of the new ES potential results from a
smooth deformation of that of the conventional SI one, we have also observed in other
examples a generation or suppression of bound states, depending on the values taken by
the mass parameters. We would like to stress the nontrivial nature of this result, of which
very few cases have been signalled in the literature devoted to PDEM SE's so far.\par
%
%
It is rather obvious that our results for bound states could be easily extended to the
$S$-matrix and that our construction method of new ES PDEM potentials could also be
applied to more complicated forms of the PDEM's or to other potentials that are SI
under parameter translation. An interesting open question for future work is whether it
could be generalized to other types of SI, such as SI under parameter scaling.\par
%
%
\section*{Acknowledgment}

AB thanks the University Grants Commission, New Delhi for the award of a Junior Research
Fellowship. CQ is a Research Director, National Fund for Scientific Research (FNRS),
Belgium.\par
%
%
\section*{Appendix}

In this appendix, we list some of the results obtained for the SI potentials considered in
table 4.1 of~\cite{cooper} when deforming the corresponding SI condition as explained in
sections 2 and 3. For simplicity's sake, the parameter dependence of the functions has
not been indicated explicitly.\par
%
%
\medskip
\noindent
{\sl Shifted oscillator}
\medskip
\[ V_{\rm eff}(x) = \frac{1}{4} \omega^2 \left(x - \frac{2b}{\omega}\right)^2 \]
\[ W(x) = \lambda x + \mu \qquad \mbox{(class 1: $\phi(x) = x$)} \]
\[ g(x) = \alpha x^2 + 2\beta x \qquad \alpha > \beta^2 \ge 0 \]
\[ \lambda = \frac{1}{2}(\alpha + \Delta) \qquad \mu = \beta -
    \frac{b\omega}{2\lambda} \qquad \Delta \equiv \sqrt{\omega^2 + \alpha^2} \]
\[ \lambda_i = \lambda + i \alpha \qquad \mu_i = \frac{\lambda\mu + 2i\beta\lambda
    + i^2\alpha\beta}{\lambda + i\alpha} \]
\begin{eqnarray*}
  E_n & = & \left(n + \frac{1}{2}\right) \Delta + \left(n^2 + n + \frac{1}{2}\right)
       \alpha + b^2 - \left(\frac{[(2n+1)\Delta + (2n^2+2n+1)\alpha] \beta -
    b\omega}{\Delta + (2n+1)\alpha}\right)^2 \nonumber \\
  &&  \quad n=0, 1, 2, \ldots  
\end{eqnarray*}
\[ \psi_0(x) \propto f^{-(\lambda+\alpha)/(2\alpha)} \exp\left(\frac{\lambda\beta -
    \mu\alpha}{\alpha\delta} \arctan \frac{\alpha x + \beta}{\delta}\right) \qquad
    \delta \equiv \sqrt{\alpha - \beta^2} \]
\[ \tv(x) = 2(\rho + 2\sigma) \alpha x (\alpha x + 2\beta) + 2\rho\alpha + 4\sigma
    \beta^2 \]
\par
%
%
\medskip
\noindent
{\sl Three-dimensional oscillator}
\medskip
\[ V_{\rm eff}(x) = \frac{1}{4} \omega^2 x^2 + \frac{l(l+1)}{x^2} \qquad 0 \le x <
\infty
\]
\[ W(x) = \frac{\lambda}{x} + \mu x \qquad \mbox{(class 2: $\phi(x) = \frac{1}{x}$)} \]
\[ g(x) = \alpha x^2 \qquad \alpha > 0 \]
\[ \lambda = - l - 1 \qquad \mu = \frac{1}{2}(\alpha + \Delta) \qquad \Delta \equiv
    \sqrt{\omega^2 + \alpha^2} \]
\[ \lambda_i = \lambda - i \qquad \mu_i = \mu + i\alpha \]
\[ E_n = \Delta \left(2n + l + \frac{3}{2}\right) + \alpha \left[2(n+l+1)(2n+1) +
\frac{1}{2}\right] \qquad n=0, 1, 2, \ldots \]
\[ \psi_0(x) \propto x^{l+1} f^{-[\mu + (l+2)\alpha]/(2\alpha)} \]
\[ \tv(x) = 2(\rho + 2\sigma) \alpha^2 x^2  + 2\rho\alpha \]
\par
%
%
\medskip
\noindent
{\sl Coulomb}
\medskip
\[ V_{\rm eff}(x) = - \frac{e^2}{x} + \frac{l(l+1)}{x^2} \qquad 0 \le x < \infty \]
\[ W(x) = \frac{\lambda}{x} + \mu \qquad \mbox{(class 1: $\phi(x) = \frac{1}{x}$)} \]
\[ g(x) = \alpha x  \qquad \alpha > 0 \]
\[ \lambda = - l - 1 \qquad \mu = - \frac{e^2 + \alpha\lambda}{2\lambda} \]
\[ \lambda_i = \lambda - i \qquad \mu_i = - \frac{e^2 + \alpha\lambda(2i+1) - \alpha
    i^2}{2(\lambda - i)} \]
\begin{eqnarray*} 
  E_n & = & - \left(\frac{e^2 - \alpha[n^2 + (l+1)(2n+1)]}{2(n+l+1)}\right)^2 \\ 
  && \quad \mbox{\rm $n=0, 1, \ldots, n_{\rm max}$, where $n_{\rm max}=$ largest
       integer such that} \\
  && \qquad \mbox{\rm $n^2 + (l+1)(2n+1) < \frac{e^2}{\alpha}$ if $\alpha <
       \frac{e^2}{l+1}$} 
\end{eqnarray*}
\[ \psi_0(x) \propto x^{l+1} f^{-\left(\frac{\mu}{\alpha} + l + \frac{3}{2}\right)} \]
\[ \tv(x) = \sigma \alpha^2 \]
\par
%
%
\medskip
\noindent
{\sl Morse}
\medskip
\[ V_{\rm eff}(x) = B^2 e^{-2x} - B (2A+1) e^{-x} \qquad A, B > 0 \]
\[ W(x) = \lambda e^{-x} + \mu \qquad \mbox{(class 1: $\phi(x) = e^{-x}$)} \]
\[ g(x) = \alpha e^{-x} \qquad \alpha > 0 \]
\[ \lambda = - \frac{1}{2}(\alpha + \Delta) \qquad \mu = - \frac{1}{2} \left(
    \frac{B(2A+1)}{\lambda} + 1\right) \qquad \Delta \equiv \sqrt{4B^2 + \alpha^2} \]
\[ \lambda_i = \lambda - i \alpha \qquad \mu_i = \frac{2\lambda(\mu-i) + i^2\alpha}
    {2(\lambda - i\alpha)} \]
\begin{eqnarray*} 
  E_n & = & - \frac{1}{4} \left(\frac{2B(2A+1) - [(2n+1)\Delta + (2n^2+2n+1)\alpha]}
       {\Delta + (2n+1)\alpha}\right)^2 \\
  && \quad \mbox{\rm $n=0, 1, \ldots, n_{\rm max}$, where $n_{\rm
       max} =$ largest integer smaller than $A$} \\  
  && \qquad \mbox{\rm and such that $\alpha < \alpha_{\rm max}(n_{\rm max})$ with}
       \\
  && \qquad \alpha_{\rm max}(0) = \frac{4A(A+1)B}{2A+1} \\
  && \qquad \alpha_{\rm max}(n) = \frac{B(2A+1)(2n^2+2n+1) - B(2n+1)
       [(2A+1)^2+4n^2(n+1)^2]^{1/2}}{2n^2(n+1)^2} \\
  && \qquad \quad n=1, 2, \ldots 
\end{eqnarray*}
\[ \psi_0(x) \propto f^{\frac{\lambda}{\alpha} - \mu - \frac{1}{2}} e^{- \mu x}\]
\[ \tv(x) = (\rho + \sigma) \alpha^2 e^{-2x} + \rho \alpha e^{-x} \]
\par
%
%
\medskip
\noindent
{\sl Eckart}
\medskip
\[ V_{\rm eff}(x) = A(A-1) \csch^2 x - 2B \coth x \qquad A \ge \frac{3}{2} \qquad
    B > A^2 \qquad 0 \le x < \infty \]
\[ W(x) = \lambda \coth x + \mu \qquad \mbox{(class 1: $\phi(x) = \coth x$)} \]
\[ g(x)= \alpha e^{-x} \sinh x \qquad  -2 \le \alpha \ne 0 \]
\[ \lambda = - A \qquad \mu = \frac{B}{A} - \frac{1}{2} \alpha \]
\[ \lambda_i = \lambda - i \qquad \mu_i = \frac{\lambda\mu - \frac{1}{2}\alpha i
    (2\lambda-i)}{\lambda - i} \]
\begin{eqnarray*}
  E_n & = & - (A+n)^2 - \left(\frac{B - \frac{1}{2}\alpha[(2n+1)A + n^2]}{A+n}\right)^2
       - \alpha[(2n+1)A + n^2] \\
  && \quad \mbox{\rm $n=0, 1, 2, \ldots \ $ if $\alpha = -2$} \\
  && \quad \mbox{\rm $n=0, 1, \ldots, n_{\rm max} \ $ if $\alpha > -2$, where
       $n_{\rm max} =$ largest integer such that} \\
  && \qquad (A+n)^2 < \frac{2B + \alpha A(A-1)}{2 + \alpha}  
\end{eqnarray*}
\begin{eqnarray*} 
  \psi_0(x) & \propto & (\coth x - 1)^{-A-1} \csch x\, \exp\left(- \frac{\mu - A}{\coth x
       - 1}\right) \qquad \mbox{\rm if $\alpha = -2$} \\
  & \propto & (\coth x + 1)^{1/2} (\coth x + 1 + \alpha)^{- \frac{(1+\alpha)A + \mu}
       {2+\alpha} - \frac{1}{2}} (\coth x - 1)^{\frac{\mu - A}{2+\alpha}} \qquad
       \mbox{\rm if $\alpha > -2$}
\end{eqnarray*}
\[ \tv(x) = (\rho + \sigma) \alpha^2 e^{-4x} - \rho \alpha (2+\alpha) e^{-2x} \]
\par
%
%
\medskip
\noindent
{\sl Scarf I}
\medskip
\[ V_{\rm eff}(x) = (B^2 + A^2 - A) \sec^2 x - B (2A-1) \tan x \sec x \qquad 0 < B <
    A-1 \qquad - \frac{\pi}{2} \le x \le \frac{\pi}{2} \]
\[ W(x) = \lambda \tan x + \mu \sec x \qquad \mbox{(class 3: $\phi(x) = \sin x$)} \]
\[ g(x) = \alpha \sin x \qquad 0 < |\alpha| < 1 \]
\[ \lambda = \frac{1}{2}(1 + \Delta_+ + \Delta_-) \qquad \mu = \frac{1}{2}(\alpha -
    \Delta_+ + \Delta_-) \qquad \Delta_{\pm} \equiv \sqrt{\frac{1}{4}(1 \mp
    \alpha)^2 + (A \pm B)(A \pm B - 1)} \]
\[ \lambda_i = \lambda + i \qquad \mu_i = \mu + i \alpha \]
\[ E_n = - \frac{1}{4}(2n + 1 + \Delta_+ + \Delta_-)^2 + \alpha \left(n + \frac{1}{2}
    \right)(\Delta_+ - \Delta_-) - \alpha^2 \left(n^2 + n + \frac{1}{2}\right) \qquad n=0,
   1, 2, \ldots  \]
\[ \psi_0(x) \propto f^{- \frac{\lambda - \alpha\mu}{1 - \alpha^2} - \frac{1}{2}} (1 -
    \sin x)^{\frac{\lambda + \mu}{2(1 + \alpha)}} (1 + \sin x)^{\frac{\lambda - \mu}{2(1
    - \alpha)}} \]
\[ \tv(x) = - (\rho + \sigma) \alpha^2 \sin^2 x - \rho\alpha \sin x + \sigma \alpha^2 \]
\par
%
%
\medskip
\noindent
{\sl Rosen-Morse I}
\medskip
\[ V_{\rm eff}(x) = A(A-1) \csc^2 x + 2B \cot x \qquad A \ge \frac{3}{2} \qquad 0 \le 
    x \le \pi \]
\[ W(x) = \lambda \cot x + \mu \qquad \mbox{(class 1: $\phi(x) = \cot x$)} \]
\[ g(x) = \sin x (\alpha\cos x + \beta\sin x) \qquad \frac{|\alpha|}{2} < \sqrt{1+\beta}
    \qquad \beta > -1 \]
\[ \lambda = - A \qquad \mu = - \frac{B}{A} - \frac{1}{2} \alpha \]
\[ \lambda_i = \lambda - i \qquad \mu_i = \frac{\lambda\mu - \frac{1}{2}\alpha i
    (2\lambda-i)}{\lambda - i} \]
\[ E_n = (A+n)^2 - \left(\frac{B + \frac{1}{2}\alpha[(2n+1)A + n^2]}{A+n}\right)^2 +
    \beta[(2n+1)A + n^2] \qquad n=0, 1, 2, \ldots  \]
\[  
  \psi_0(x) \propto f^{-(A+1)/2} (\sin x)^A \exp\left(\frac{\mu + \frac{1}{2}
       \alpha A}{\delta} \arctan \frac{\cot x + \frac{\alpha}{2}}{\delta}\right) \qquad
       \delta \equiv \sqrt{1 + \beta - \frac{\alpha^2}{4}} \]
\begin{eqnarray*}
  \tv(x) & = & (\rho + \sigma) \left[\frac{1}{2}(\alpha^2 - \beta^2)\cos 4x +
       \alpha\beta \sin 4x\right] + \rho (2+\beta) (- \alpha\sin 2x + \beta\cos 2x) \\
  && \mbox{} + (- \rho + \sigma) \frac{1}{2}(\alpha^2 + \beta^2)
\end{eqnarray*}
%
%


\newpage
\begin{thebibliography}{99}

\bibitem{geller} Geller M R and Kohn W 1993 {\sl Phys.\ Rev.\ Lett.} {\bf 70} 3103

\bibitem{serra} Serra Ll and Lipparini E 1997 {\sl Europhys.\ Lett.} {\bf 40} 667

\bibitem{barranco} Barranco M, Pi M, Gatica S M, Hern\'andez E S and Navarro J 1997
{\sl Phys.\ Rev.} B {\bf 56} 8997

\bibitem{ring} Ring P and Schuck P 1980 {\sl The Nuclear Many Body Problem}
(New York: Springer)

\bibitem{arias} Arias de Saavedra F, Boronat J, Polls A and Fabrocini A 1994 {\sl Phys.\
Rev.} B {\bf 50} 4248

\bibitem{puente} Puente A, Serra Ll and Casas M 1994 {\sl Z.\ Phys.} D {\bf 31} 283

\bibitem{levy} L\'evy-Leblond J-M 1995 {\sl Phys.\ Rev.} A {\bf 52} 1845

\bibitem{chetouani} Chetouani L, Dekar L and Hammann T F 1995 {\sl Phys.\ Rev.} A
{\bf 52} 82

\bibitem{yung} Yung K C and Yee J H 1994 {\sl Phys.\ Rev.} A {\bf 50} 104

\bibitem{dekar} Dekar L, Chetouani L and Hammann T F 1998 {\sl J.\ Math.\ Phys.} {\bf 39}
2551\\
Dekar L, Chetouani L and Hammann T F 1999 {\sl Phys.\ Rev.} A {\bf 59} 107

\bibitem{milanovic} Milanovi\'c V and Ikoni\'c Z 1999 {\sl J.\ Phys.\ A: Math.\ Gen.} {\bf 32}
7001

\bibitem{plastino} Plastino A R, Rigo A, Casas M, Garcias F and Plastino A 1999 {\sl Phys.\
Rev.} A {\bf 60} 4318\\
Plastino A R, Puente A, Casas M, Garcias F and Plastino A 2000 {\sl Rev.\ Mex.\ Fis.} {\bf 46}
78

\bibitem{dutra} de Souza Dutra A and Almeida C A S 2000 {\sl Phys.\ Lett.} A {\bf 275}
25 \\
de Souza Dutra A, Hott M and Almeida C A S 2003 {\sl Europhys.\ Lett.} {\bf 62} 8

\bibitem{roy} Roy B and Roy P 2002 {\sl J.\ Phys.\ A: Math.\ Gen.} {\bf 35} 3961

\bibitem{koc} Ko\c c R, Koca M and K\"orc\"uk E 2002 {\sl J.\ Phys.\ A: Math.\ Gen.} {\bf 35}
L527\\
Ko\c c R and Koca M 2003 {\sl J.\ Phys.\ A: Math.\ Gen.} {\bf 36} 8105

\bibitem{alhaidari} Alhaidari A D 2002 {\sl Phys.\ Rev.} A {\bf 66} 042116

\bibitem{gonul} G\"on\"ul B, G\"on\"ul B, Tutcu D and \"Ozer O 2002 {\sl Mod.\ Phys.\
Lett.} A {\bf 17} 2057\\
G\"on\"ul B, \"Ozer O, G\"on\"ul B and \"Uzg\"un F  2002 {\sl Mod.\ Phys.\ Lett.} A {\bf 17}
2453

\bibitem{cq04} Quesne C and Tkachuk V M 2004 {\sl J.\ Phys.\ A: Math.\ Gen.} {\bf 37}
4267

\bibitem{bagchi} Bagchi B, Gorain P, Quesne C and Roychoudhury R 2004 {\sl Mod.\
Phys.\ Lett.} A {\bf 19} 2765
\\ Bagchi B, Gorain P, Quesne C and Roychoudhury R 2004 {\sl Czech.\ J.\ Phys.} {\bf 54}
1019

\bibitem{yu} Yu J, Dong S-H and Sun G-H 2004 {\sl Phys.\ Lett.} A {\bf 322} 290 \\
Yu J and Dong S-H 2004 {\sl Phys.\ Lett.} A {\bf 325} 194

\bibitem{bhatta} Bhattacharjie A and Sudarshan E C G 1962 {\sl Nuovo Cimento} {\bf 25}
864\\
Natanzon G A 1979 {\sl Theor.\ Math.\ Phys.} {\bf 38} 146\\
L\'evai G 1989 {\sl J.\ Phys.\ A: Math.\ Gen.} {\bf 22} 689

\bibitem{alhassid} Alhassid Y, G\" ursey F and Iachello F 1986 {\sl Ann.\ Phys., N.Y.} {\bf
167} 181\\
Wu J and Alhassid Y 1990 {\sl J.\ Math.\ Phys.} {\bf 31} 557\\
Englefield M J and Quesne C 1991 {\sl J.\ Phys.\ A: Math.\ Gen.} {\bf 24} 3557\\
L\'evai G 1994 {\sl J.\ Phys.\ A: Math.\ Gen.} {\bf 27} 3809

\bibitem{gendenshtein} Gendenshtein L E 1983 {\sl JETP Lett.} {\bf 38} 356\\
Dabrowska J, Khare A and Sukhatme U 1988 {\sl J.\ Phys.\ A: Math.\ Gen.} {\bf 21} L195

\bibitem{cooper} Cooper F, Khare A and Sukhatme U 1995 {\sl Phys.\ Rep.} {\bf 251}
267

\bibitem{mizrahi} Mizrahi S S, Camargo Lima J P and Dodonov V V 2004 {\sl J.\ Phys.\ A:
Math.\ Gen.} {\bf 37} 3707

\bibitem{spiridonov} Spiridonov V 1992 {\sl Phys.\ Rev.\ Lett.} {\bf 69} 398 \\
Spiridonov V 1992 {\sl Mod.\ Phys.\ Lett.} A {\bf 7} 1241

\bibitem{khare} Khare A and Sukhatme U P 1993 {\sl J.\ Phys.\ A: Math.\ Gen.} {\bf 26}
L901 \\
Barclay D T, Dutt R, Gangopadhyaya A, Khare A, Pagnamenta A and Sukhatme U 1993
{\sl Phys.\ Rev.} A {\bf 48} 2786

\bibitem{sukhatme} Sukhatme U P, Rasinariu C and Khare A 1997 {\sl Phys.\ Lett.} A
{\bf 234} 401 \\
Gangopadhyaya A, Mallow J V, Rasinariu C and Sukhatme U P 1999 {\sl Theor.\ Math.\
Phys.} {\bf 118} 285

\bibitem{loutsenko} Loutsenko I, Spiridonov V, Vinet L and Zhedanov A 1998 {\sl J.\ Phys.\
A: Math.\ Gen.} {\bf 31} 9081

\bibitem{morrow} Morrow R A 1987 {\sl Phys.\ Rev.} B {\bf 35} 8074

\bibitem{ribeiro} Ribeiro Filho J, Farias G A and Freire V N 1996 {\sl Braz.\ J.\ Phys.}
{\bf 26} 388

\bibitem{cavalcante} Cavalcante F S A, Costa Filho R N, Ribeiro Filho J, de
Almeida C A S and Freire V N 1997 {\sl Phys.\ Rev.} B {\bf 55} 1326

\bibitem{vonroos} von Roos O 1983 {\sl Phys.\ Rev.} B {\bf 27} 7547

\bibitem{bendaniel} BenDaniel D J and Duke C B 1966 {\sl Phys.\ Rev.} B {\bf 152} 683

\bibitem{bastard} Bastard G 1981 {\sl Phys.\ Rev.} B {\bf 24} 5693

\bibitem{zhu} Zhu Q-G and Kroemer H 1983 {\sl Phys.\ Rev.} B {\bf 27} 3519 

\bibitem{li} Li T L and Kuhn K J 1993 {\sl Phys.\ Rev.} B {\bf 47} 12760

\bibitem{sukumar85b} Sukumar C V 1985 {\sl J.\ Phys.\ A: Math.\ Gen.} {\bf 18} L57

\bibitem{cq99} Quesne C 1999 {\sl J.\ Phys.\ A: Math.\ Gen.} {\bf 32} 6705

\bibitem{sukumar85a} Sukumar C V 1985 {\sl J.\ Phys.\ A: Math.\ Gen.} {\bf 18} 2917

\bibitem{kwong} Kwong W and Rosner J L 1986 {\sl Prog.\ Theor.\ Phys.\ Suppl.} {\bf
86} 366 \\
Bagchi B 1990 {\sl Int.\ J.\ Mod.\ Phys.} A {\bf 5} 1763

\bibitem{nieto} Nieto M M 1978 {\sl Phys.\ Rev.} A {\bf 17} 1273

\bibitem{davies} Davies P C W 1984 {\sl Quantum Mechanics} (London: Routledge and
Kegan Paul)

\end {thebibliography}
\end{document}